# Exploring Artificial Intelligence and Culture: Methodology for a comparative study of AI's impact on norms, trust, and problem-solving across academic and business environments

Matthias Hümmer 1*; Theophile Shyiramunda 2*; Michelle J. Cummings-Koether 3*

**1* Prof. Dr.-Ing. Matthias Hümmer,**

Professor, Institute for the Transformation of Society (I-ETOS), Deggendorf Institute of Technology - European Campus Rottal-Inn

E-Mail: matthias.huemmer@th-deg.de

ORCID ID: https://orcid.org/0009-0003-8122-470X

**2* Dr. Theophile Shyiramunda,**

Research Associate, Institute for the Transformation of Society (I-ETOS), Deggendorf Institute of Technology - European Campus Rottal-Inn

E-Mail: theophile.shyiramunda@th-deg.de

ORCID ID: https://orcid.org/0000-0001-6725-3756

**3* Prof. Dr. Michelle J. Cummings-Koether,**

Professor, Institute for the Transformation of Society (I-ETOS), Deggendorf Institute of Technology - European Campus Rottal-Inn

E-Mail: michelle.cummings-koether@th-deg.de

ORCID ID: https://orcid.org/0000-0002-7137-3539

**Corresponding author @ Prof. Dr.-Ing. Matthias Hümmer,** Professor, Institute for the Transformation of Society (I-ETOS), Technische Hochschule Deggendorf - European Campus Rottal-Inn, E-Mail: matthias.huemmer@th-deg.de, ORCID ID: https://orcid.org/0009-0003-8122-470X

**Abstract**

This paper presents a rigorous methodological framework to examine the bidirectional relationship between artificial intelligence (AI), human cognition, problem-solving, and cultural adaptation across academic and business environments. Addressing a critical gap in current research, the study investigates how AI shapes cognitive processes and organizational norms, while cultural values and institutional contexts influence AI adoption, trust formation, and usage behaviours over time. The research employs a three-wave longitudinal design over five months, tracking participants' AI knowledge, perceived competence, trust development, and cultural responses. Participants from academic institutions, including the European Campus Rottal-Inn (ECRI) and diverse business organizations provide a comparative perspective on context-specific AI integration. A dynamic participant pool, reflecting continuous, intermittent, and wave-specific respondents, mirrors real-world organizational variability, enhancing ecological validity. Methodologically, the study combines quantitative longitudinal modelling with qualitative thematic analysis, capturing temporal, structural, and cultural patterns in AI adoption. It traces AI acculturation through phases of initial resistance, exploratory adoption, and cultural embedding, revealing differences in trust trajectories and problem-solving strategies across organizational contexts. Academic settings emphasize collaborative, deliberative integration, whereas business contexts prioritize performance-driven adoption. By framing AI adoption as bidirectional, the framework challenges deterministic models, showing that AI both reflects and reshapes organizational norms, decision-making, and cognitive engagement. As the first longitudinal comparative study of its kind, this research advances methodological rigor and provides a foundation for human-centred, culturally responsive AI integration strategies.

**Keywords:** artificial intelligence adoption, organizational culture, organizational change, technology acceptance, cultural adaptation, trust formation

**Statements and Declarations**

**Competing Interests**

All authors certify that they have no affiliations with or involvement in any organization or entity with any financial interest or non-financial interest in the subject matter or materials discussed in this manuscript. On behalf of all authors, the corresponding author states that there is no conflict of interest.

**Data availability**

This study is methodological in nature and does not employ empirical data. Instead, example datasets were synthetically generated solely for the purpose of illustrating the proposed methodological procedures. These example scripts and datasets are available from the corresponding author upon reasonable request.

**Author contributions**

**Funding**

The authors did not receive support from any organization for the submitted work.

**Introduction**

The accelerating evolution and widespread integration of Artificial Intelligence (AI), particularly Large Language Models (LLMs), has reshaped the ways in which individuals learn, work, and interact across societal, academic, and business contexts. These systems not only democratize access to knowledge and augment problem-solving capacities (Brown et al., 2020; Peláez-Sánchez et al., 2024), but they also exert profound effects on human cognition, trust dynamics, and cultural practices (Ahmad et al., 2023; Gerlich, 2025; Wang & Fan, 2025). Recent evidence underscores a paradox: while AI can enhance efficiency and decision-making (Hancock et al., 2021; Sajja et al., 2024), over-reliance fosters "cognitive debt," diminishing deep analytical reasoning, independent problem-solving, and creativity (Kosmyna et al., 2024; Stadler et al., 2024). Longitudinal studies suggest that prolonged AI dependence risks cognitive atrophy, as illustrated by reduced neural activity and homogenized expression in AI-assisted writing (Kosmyna et al., 2025; Jelson et al., 2025), alongside empirical findings of de-skilling in clinical contexts when AI assistance is removed. For instance, a recent observational study reported in *TIME* found that within six months of using AI for polyp detection, clinicians became over-reliant, leading to decreased performance when AI assistance was removed, manifesting as being "less motivated, less focused, and less responsible when making cognitive decisions without AI assistance" (Jeyaretnam, 2025). Beyond individual cognition, AI integration increasingly mediates cultural practices, organizational trust, and institutional norms, demanding careful analysis of how technology both mirrors and transforms the values of the communities in which it operates (Nisbett, 2003; Milana et al., 2024; Yang et al., 2025).

Against this backdrop, our methodological paper introduces a comparative, longitudinal framework to investigate the bidirectional relationship between AI and cultural adaptation within academic and business environments. We focus on how AI technologies influence cognitive processes, trust formation, and problem-solving behaviours, while also exploring how cultural norms and organizational values condition AI adoption and integration (Benbya et al., 2020; Scott, 2014; Atf & Lewis, 2025; Gillespie et al., 2025). Specifically, this paper situates our work within three thematic domains: first, we examine the impact of AI on cognitive processes and cultural dynamics; second, we outline the objectives of our study; and third, we preview the paper's structure. Following this introduction, we present a synthesis of related works, identify critical research gaps, and discuss the relevance of our approach. We then detail our research methodology, highlight methodological considerations and limitations, and conclude by underscoring the expected contributions and implications of our study for responsible, culturally responsive AI integration strategies.

**The impact of AI on Cognitive Processes and Cultural Dynamics**

Emerging research underscores a dualistic influence of AI technologies on human cognitive functioning. While these systems enhance immediate problem-solving capabilities and operational efficiency (Hancock et al., 2021), they simultaneously risk diminishing critical cognitive functions such as deep analytical processing and autonomous problem-solving abilities. This phenomenon, recently termed "cognitive debt" describes the accumulation of reduced cognitive engagement that occurs when individuals rely heavily on AI assistance for complex tasks (Kosmyna et al., 2024). This tension highlights the critical importance of examining how AI usage varies across different cultural and organizational contexts and how it influences human behaviour over time.

Recent empirical evidence deepens this concern. A 2025 MIT study found that students using ChatGPT to write essays showed significantly reduced neural engagement, as measured by EEG, alongside diminished memory and creativity, compared to peers using only their own thinking or search engines. Their performance declined over repeated tasks, suggesting emerging cognitive atrophy in AI-dependent users (Kosmyna et al., 2025). Similarly, a Time-Lancet–reported clinical study observed that clinicians who relied on AI support in colonoscopy saw their diagnostic skills decline when AI assistance was removed, illustrating a real-world "Google Maps effect" of de-skilling (2025). At the same time, studies have explored effects on cultural and cognitive diversity. According to Vashistha (2025), AI-assisted writing tends to homogenize expression, essays converge on similar themes and language, which undermines individual voice and may erode cultural diversity in discourse over time

The integration of artificial intelligence technologies across organizational landscapes has fundamentally transformed approaches to problem-solving, decision-making, and cultural adaptation (Russell & Norvig, 2021; Brynjolfsson & McAfee, 2014). While AI represents one of the most significant technological advances of the 21st century, its adoption and integration are deeply intertwined with cultural contexts, organizational values, and institutional frameworks (Nisbett, 2003; Hall, 1976). This intersection between technological innovation and cultural dynamics generates a complex research landscape that demands both sophisticated methodological approaches and longitudinal investigation.

A particularly underexplored area concerns the temporal dimension of AI integration. Most existing studies employ cross-sectional designs, capturing attitudes and behaviours at single time points while failing to account for the dynamic nature of cultural adaptation (Benbya et al., 2020; Marangunić & Granić, 2015). This methodological limitation prevents comprehensive understanding of how initial resistance evolves into acceptance, how trust develops through experience, and how cultural norms shift with ongoing exposure to AI technologies. The present study addresses this critical gap through a longitudinal methodology designed to capture the dynamic interplay between AI technologies and cultural contexts over time.

The theoretical foundations for this research draw from multiple disciplinary perspectives, including cultural adaptation theory, technology acceptance models, and organizational behaviour frameworks. The Technology Acceptance Model (TAM) and its extensions offer robust frameworks for understanding individual-level adoption decisions, with recent work incorporating ethical considerations, trust mechanisms, and cultural moderators into predictive models (Davis, 1989; Marangunić & Granić, 2015). However, these models require extension to account for the bidirectional relationship between technology and culture, wherein AI systems both respond to and actively reshape cultural contexts through sustained interaction.

The significance of this research extends beyond theoretical contributions to practical implications for AI implementation, policy development, and organizational change management. As AI becomes increasingly sophisticated and ubiquitous, understanding how various cultural contexts facilitate or constrain AI adoption becomes essential for successful implementation. Organizations investing in AI technologies require evidence-based guidance to navigate cultural adaptation challenges and develop integration strategies aligned with institutional values and stakeholder expectations.

This study specifically addresses three critical gaps in existing literature. First, the lack of comparative research examining AI adoption across different organizational cultures limits and understanding of context-dependent integration factors. Second, the absence of longitudinal research designs prevents comprehensive insight into temporal patterns of trust development and cultural adaptation processes. Third, limited attention has been paid to the bidirectional relationship between AI technologies and cultural norms, as most studies focus on how culture shapes technology adoption rather than examining how technology transforms cultural practices over time.

**Research Objectives**

This study investigates the integration and impact of artificial intelligence (AI) within academic and business contexts through a longitudinal, comparative framework. Its objectives are to characterize self-perceived AI competence across organizational contexts, elucidating how individuals evaluate their AI knowledge and capabilities; analyse cultural and organizational determinants of AI adoption, revealing how contextual norms shape AI use and integration; examine AI's influence on problem-solving and cognitive processes, capturing how sustained engagement alters decision-making and cognitive functioning; and explore the bidirectional interplay between AI and cultural practices, assessing how repeated AI interaction transforms social behaviours, trust, and organizational norms.

Additionally, the study evaluates the reciprocal relationship between AI technologies and cultural practices, assessing whether AI systems modify cultural norms, trust dynamics, and social interactions over the course of repeated exposure. The longitudinal design allows for tracking potential shifts in human behavior patterns, enabling detailed understanding of how prolonged AI engagement might recalibrate individual and collective cognitive functions and social practices (Buolamwini & Gebru, 2018; Floridi, 2020). By addressing these objectives, the study offers theoretical insights into technology-culture dynamics and actionable guidance for leaders, policymakers, and AI developers, establishing a robust framework for understanding the evolving landscape of human-AI interaction.

Related Work

Artificial intelligence (AI) is fundamentally transforming how individuals and organizations approach problem-solving across various socio-cultural and institutional contexts. While AI technologies promise increased efficiency, accuracy, and innovation, their deployment and acceptance are not culturally neutral. Rather, they are both shaped by and actively shape the cultural milieus in which they are embedded (Peláez-Sánchez et al., 2024; Dwivedi et al., 2021; Glikson & Woolley, 2020). In recognizing that technology is not merely a technical artifact but a socio-cultural construct, it becomes imperative to investigate the mutual influence between AI and cultural dimensions in problem-solving. This literature review synthesizes existing scholarly work in four core areas (see Table 1): (1) cultural influences on AI use in problem-solving, (2) AI's feedback effects on cultural norms and practices, and (3) AI impact on human on problem-solving abilities; (4) research gaps relevant to the present study's design.

Table 1. Literature review core area and description

| Core Area | Description |
| --- | --- |
| Cultural influences on AI use in problem-solving | Explores how cultural values, norms, and institutional frameworks shape the adoption and application of AI in solving problems. |
| AI's feedback effects on cultural norms and practices | Examines how AI technologies reshape cultural practices, communication styles, and organizational values through feedback loops. |
| AI impact on human problem-solving Abilities | Analyses cognitive and behavioural shifts in human problem-solving capabilities when tasks are mediated or augmented by AI systems. |
| Research gaps relevant to the present study's design | Identifies areas lacking sufficient investigation, such as longitudinal effects, comparative cultural studies, and bidirectional dynamics. |

Source: Authors 'own Illustration

**AI and Cultural Dimensions in Problem-Solving**

Artificial intelligence (AI) is rapidly altering the ways in which both individuals and organizations engage with complex problem-solving. However, the adoption and acceptance of these technologies are not universal; they are filtered through cultural lenses. Scholars such as Glikson and Woolley (2020) and Nisbett (2003) have shown that AI cannot be understood as a neutral instrument but rather as a set of tools that mirror, and in turn reshape, the cognitive patterns, social practices, and values of the communities where they are implemented. This perspective highlights the need to view technology as more than a technical product. It should also be analysed as a sociocultural construct. Culture, understood as a network of shared norms, values, and expectations that guide social conduct and mental processes (Hofstede, Hofstede, & Minkov, 2010), plays a decisive role in shaping the trajectory of technological adoption. As Reimer et al. (2014) argue, cultural frameworks strongly influence decision-making preferences, while more recent studies (Milana et al., 2024) demonstrate that these frameworks also condition how individuals evaluate the risks and benefits of AI in diverse organizational settings.

Among the most influential theoretical models is Hofstede's framework, which highlights dimensions such as uncertainty avoidance, individualism versus collectivism, and power distance (Hofstede et al., 2010). Cultures high in uncertainty avoidance are more likely to resist adopting opaque or unpredictable technologies, including AI, and demand greater transparency and validation (Zhou et al., 2024). Conversely, cultures low in uncertainty avoidance tend to be more tolerant of ambiguity and open to innovative AI applications (Glikson & Woolley, 2020). Schwartz's theory of basic human values also sheds light on these dynamics, particularly through the dimensions of hierarchy versus egalitarianism and embeddedness versus autonomy, which can influence how societies evaluate and accept AI (Schwartz, 2012). Similarly, institutional and organizational theory suggests that organizations embedded in different cultural contexts approach AI adoption in ways that reflect broader social norms and expectations (Scott, 2014; Su & Yang, 2023).

The individualism collectivism dimension plays a critical role. Collectivist societies may hesitate to rely on AI if it is seen as undermining interpersonal trust or established hierarchies, while individualist cultures tend to frame AI as a tool for personal empowerment and efficiency (Shin, 2021; Yang et al., 2025). Power distance, the degree to which hierarchy is accepted, also shapes AI acceptance: in high power-distance cultures, AI may be more readily adopted if endorsed by authority figures, whereas low power-distance contexts demand more accountability and transparency (Hofstede et al., 2010; Bai et al., 2023). Large-scale cross-national surveys and comparative studies have supported these observations. For example, Zhang et al. (2024) found that academic self-efficacy and stress interact with cultural expectations to influence AI usage behavior in education. Similarly, Fan et al. (2024) reported that generative AI can promote efficiency but also foster metacognitive laziness, which may be interpreted differently across cultural settings depending on norms regarding effort and autonomy.

Cross-cultural psychology further demonstrates how differences in thinking styles, such as the distinction between analytic (Western) and holistic (Eastern) cognition, impact perceptions and acceptance of AI. Nisbett (2003) noted that holistic thinkers tend to evaluate technology in context, considering social and ethical implications, while analytic thinkers focus on performance and outcomes. Institutional theory provides additional context: Scott's (2014) framework suggests that acceptance of AI is shaped not only by technical performance but also by its alignment with cultural-cognitive and normative expectations. In sum, AI adoption and its use in problem-solving are tightly interwoven with cultural variables. The existing literature points to the importance of understanding these dynamics to ensure successful and responsible AI integration.

**The Influence of AI on Cultural Norms**

While culture plays a significant role in shaping the adoption and use of AI, the integration of AI technologies can also drive substantial changes in cultural norms, practices, and institutional routines. The relationship between AI and culture is bidirectional: as organizations and societies adapt to new technologies, AI itself becomes an agent of cultural transformation.

In educational settings, the introduction of AI-driven platforms has begun to alter classroom dynamics, learning behaviors, and the roles of teachers and students. Su and Yang (2023) propose a framework for implementing generative AI in education that emphasizes learner agency and context-sensitive personalization, challenging traditional hierarchies of knowledge and authority. Similarly, Milana et al. (2024) emphasizes the need for adult education to respond to AI's pedagogical disruption by rethinking the role of the educator and shifting toward dialogic, learner-centered approaches. Beyond education, organizational research has demonstrated that AI can prompt significant shifts in collaboration, communication, and decision-making processes. Faraj, Pachidi, and Sayegh (2018) argue that AI introduces new forms of distributed agency and can alter boundaries of expertise, requiring individuals and teams to develop new competencies such as algorithmic literacy and collaborative sense-making. Recent studies by Benbya, Nan, Tanriverdi, and Yoo (2020) and by Gerlich (2025) further show that AI-driven automation encourages cognitive offloading, reshaping expectations around critical thinking and information ownership in professional settings.

Empirical studies have documented these transformative effects. For example, Sajja et al. (2024) describes how personalized AI learning assistants in higher education altered student engagement patterns, decision-making behavior, and perceptions of knowledge authority. Likewise, Wang and Fan (2025), in a meta-analysis, found that students exposed to generative AI tools reported increased efficiency but also exhibited reduced higher-order thinking, suggesting that technological integration can recalibrate cognitive and pedagogical norms. AI's influence on culture also manifests in subtle changes to language and organizational discourse. King (2023) and Thompson and García (2022) observed shifts in metaphors and discourse surrounding AI-mediated writing and assessment, reflecting broader epistemic changes in how institutions define competence, authorship, and creativity.

At the institutional level, Scott's (2014) theory of institutional pillars suggests that successful AI adoption is often accompanied by changes in regulative, normative, and cultural-cognitive frameworks. This dynamic is echoed in recent educational literature, where authors such as Peláez-Sánchez et al. (2024) and Cacicio and Riggs (2023) highlight how the institutional embrace of AI compels schools and universities to renegotiate roles, responsibilities, and pedagogical values. Nissenbaum's (2001) concept of "contextual integrity" remains relevant here, as the integration of AI technologies often provokes explicit conversations about values and ethical priorities that were previously implicit. The integration of AI technologies is a dynamic process that does not merely reflect existing cultural norms but also acts as a catalyst for cultural adaptation. As organizations and institutions increasingly incorporate AI into their practices, they often experience a reconfiguration of values, norms, and processes, demonstrating that technology and culture evolve together in complex, mutually influential ways.

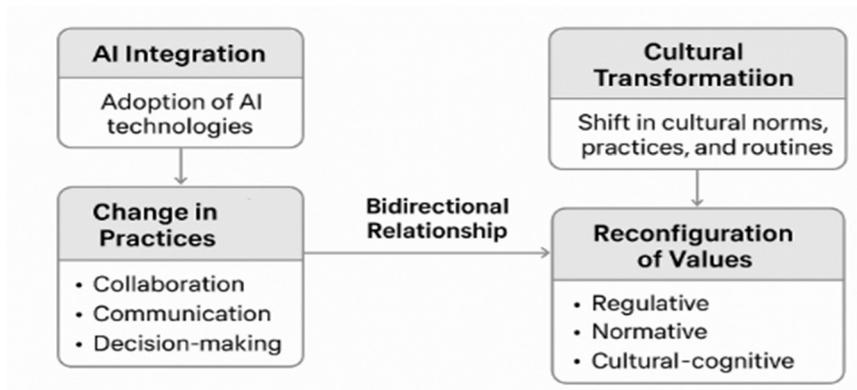

Figure 1: Influence of AI on culture

Source: Authors own illustration

Figure1 illustrates the bidirectional relationship between artificial intelligence (AI) integration and cultural transformation in educational and organizational contexts. On the left, AI integration initiates changes in everyday practices, particularly in collaboration, communication, and decision-making. These practice-level shifts are not merely operational but gradually reshape how individuals and groups engage with knowledge, authority, and problem-solving. On the right, these changes accumulate into broader cultural transformations, altering norms, routines, and institutional logics. This transformation often requires a reconfiguration of values across Scott's (2014) three institutional pillars: regulative, normative, and cultural-cognitive. The central bidirectional arrows emphasize the mutual influence of the two domains: while culture conditions how AI is adopted and used, AI also acts as a catalyst for cultural adaptation, prompting new forms of trust, new expectations of cognitive engagement, and the renegotiation of institutional roles. In this way, the framework highlights how AI and culture evolve together in a dynamic, reciprocal process rather than in a one-directional manner.

### AI impact on human on problem-solving abilities

Large Language Models (LLMs) have profoundly impacted various aspects of daily life, influencing how individuals learn, work, and engage within societal and organizational contexts. These advanced AI systems have enhanced personalized learning experiences and democratized access to information (Brown et al., 2020; Peláez-Sánchez et al., 2024), significantly affecting cognitive processes, trust dynamics, and cultural practices in both academic and business environments (Ahmad et al., 2023; Russell & Norvig, 2021; Gerlich, 2025).

Research underscores a complex duality in AI technologies: while enhancing immediate problem-solving efficiency and task performance (Hancock et al., 2021; Stadler, Bannert, & Sailer, 2024), they concurrently risk impairing essential cognitive functions, such as deep analytical processing and autonomous problem-solving capabilities (Gerlich, 2025; Ahmad et al., 2023; Fan et al., 2024; Bai, Liu, & Su, 2023). This duality emphasizes the importance of exploring variations in AI usage across diverse cultural and organizational contexts, particularly focusing on how these differences evolve and affect human behavior over time.

Cognitive offloading, the delegation of cognitive tasks to AI systems, is one significant phenomenon documented in recent literature, revealing a negative correlation between frequent AI usage and critical thinking skills (Gerlich, 2025; Risko & Gilbert, 2016). The "*use it or lose it*" neurobiological principle further supports concerns regarding potential cognitive deterioration resulting from excessive AI dependence (Shors et al., 2012; Fan et al., 2024). Automation bias, where individuals unquestioningly accept AI-generated recommendations, further exemplifies potential pitfalls of AI over-reliance, impacting decision-making quality (Parasuraman & Manzey, 2010; Skitka, Mosier, & Burdick, 2000; Lee et al., 2025).

Age-related differences also moderate AI dependency, with younger users exhibiting higher reliance on AI tools and lower critical thinking scores compared to older users, raising concerns about developmental impacts (Gerlich, 2025). Neurological research suggests that extensive use of digital technologies, including AI, may alter brain structures such as reduced gray matter density in the frontal cortex, potentially impairing cognitive functions like decision-making and analytical thinking (Loh & Kanai, 2016). Emerging neurocognitive studies have begun exploring how intensive AI interaction affects brain activity during information-seeking and memory-related tasks, with findings suggesting measurable shifts in cognitive engagement and attention regulation (Sparrow, Liu, & Wegner, 2011; Dong & Potenza, 2016).

Educational settings illustrate both the risks and benefits of AI integration. Cognitive Load Theory suggests that effective AI integration must balance cognitive support with active cognitive engagement to foster higher-order thinking, aligning with Bloom's Taxonomy of educational objectives (Sweller, van Merrienboer, & Paas, 2019; Sundararajan & Adesope, 2020; Anderson & Krathwohl, 2001). Yet, current research identifies significant risks, notably the erosion of human decision-making capabilities through excessive AI dependence (Ahmad et al., 2023; Zhang et al., 2024).

Despite these concerns, evidence also suggests positive cognitive effects of AI. AI systems can augment human capabilities when designed to support, rather than replace, human cognitive processes (Brynjolfsson & McAfee, 2014; Russell & Norvig, 2021; Sajja et al., 2024). Effective cognitive offloading, where AI takes on routine cognitive tasks, can potentially free human cognitive resources for higher-level thinking and creativity (Risko & Gilbert, 2016; Bai et al., 2023).

Finally, the literature highlights several moderating factors influencing AI's cognitive impact, including educational background, professional experience, cultural context, and individual differences in metacognitive awareness (Marangunić & Granić, 2015; Benbya et al., 2020; Kasneci et al., 2023). Addressing these moderators is crucial for developing strategies that maximize AI's cognitive benefits while mitigating potential adverse effects. Thus, this study offers a comprehensive methodological framework designed to assess AI's multifaceted impact on cognitive processes, problem-solving capabilities, and cultural norms. By capturing these dynamics over time, the research aims to bridge critical gaps in understanding human-AI interactions across diverse academic and professional contexts.

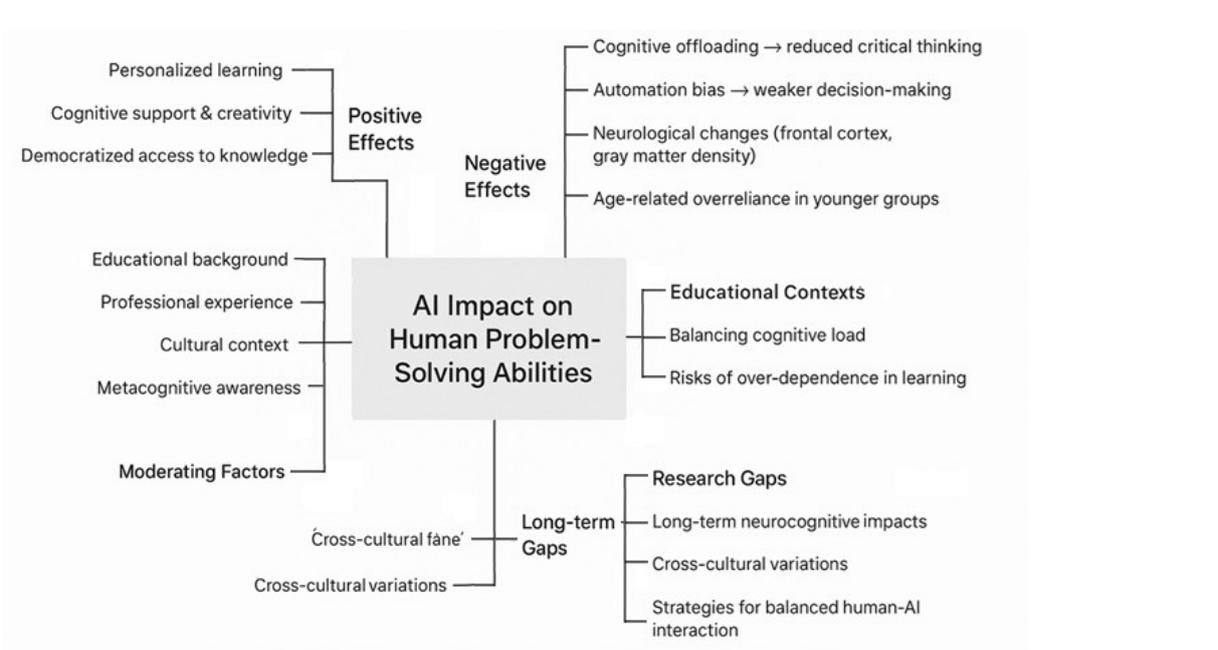

Figure 2: AI impact on Human Problem-Solving abilities

Source: Authors own illustration

As it is depicted in Figure 2, the multifaceted impact of AI on human problem-solving abilities, highlighting both positive and negative effects. On the positive side, AI enhances personalized learning, supports cognitive processes and creativity, and democratizes access to knowledge, thereby improving efficiency and skill development. Conversely, overreliance on AI can lead to cognitive offloading, reduced critical thinking, automation bias, and even neurological changes, with younger users particularly susceptible to dependency. The diagram also identifies key moderating factors such as educational background, professional experience, cultural context, and metacognitive awareness, which influence how individuals engage with AI. In educational contexts, careful integration of AI is needed to balance cognitive load and prevent over-dependence while fostering higher-order thinking. Finally, research gaps, including long-term neurocognitive impacts, cross-cultural variations, and strategies for balanced human-AI interaction, emphasize the ongoing need to study and optimize AI's role in enhancing problem-solving without undermining human cognition.

**Research Gaps and Relevance**

Despite a growing body of research on the interplay between AI and cultural dynamics, important gaps persist in the existing literature. Many studies treat culture as a static set of variables, rather than as dynamic and evolving influences that interact with technology over time (Benbya et al., 2020). This static perspective limits our understanding of how cultural perceptions, values, and practices shift in response to repeated or sustained exposure to AI. Recent work by Peláez-Sánchez et al. (2024) and Kasneci et al. (2023) echoes this concern, calling for more nuanced investigations into the reciprocal evolution of AI and sociocultural contexts, particularly within education and knowledge work.

A significant methodological limitation is the predominance of cross-sectional studies, which capture only a snapshot of attitudes or behaviors and overlook the processes by which trust, acceptance, and adaptation develop over time (Marangunić & Granić, 2015; Benbya et al., 2020). As a result, little is known about how initial skepticism may give way to acceptance, or how organizations and individuals recalibrate their expectations and practices as they gain experience with AI systems. Prahl and Van Swol (2021) emphasize the importance of longitudinal research in uncovering the evolution of trust and collaboration with AI across repeated interactions. Recent empirical studies, such as those by Fan et al. (2024) and Jelson et al. (2025), underscore how prolonged exposure to generative AI tools can reshape not only problem-solving behavior but also user perceptions of effort, authorship, and trust.

Comparative research that explores how demographic factors such as age, education, professional identity, and language shape trust in and use of AI across different contexts also remains limited. Zhang and Dafoe (2019) and Schepman and Rodway (2020) found that such variables can significantly moderate trust formation and problem-solving with AI, particularly as problem complexity increases. Gerlich (2025) adds to this line of inquiry by identifying age-dependent differences in AI reliance and critical thinking, raising questions about how generational cognitive patterns evolve in AI-mediated environments.

Another gap lies in the narrow focus of many studies on how culture shapes technology adoption, rather than considering the reciprocal influence of technology on institutional values and organizational practices. Shneiderman (2020) advocates for a human-centered approach to AI design and implementation that prioritizes transparency, trust, and the preservation of social values. Susskind (2020) highlights the ethical and cultural implications of AI increasingly mediating or even replacing, human decision-making, underscoring the need for research on adaptive strategies that ensure meaningful human oversight and the alignment of AI systems with core institutional and societal values. In support, Shen et al. (2023) and Milana et al. (2024) emphasize that LLMs and conversational agents are already shifting authority structures and ethical expectations in education and professional settings.

Also, research on the long-term cognitive effects of AI usage, particularly regarding the "use it or lose it" principle of cognitive development, remains limited. While emerging research has identified concerning patterns of cognitive offloading and potential degradation of critical thinking skills, longitudinal studies tracking cognitive development over extended periods of AI usage are scarce. Stadler, Bannert, and Sailer (2024) and Bai, Liu, and Su (2023) provide early evidence of reduced mental effort and inquiry depth among student users of LLMs, but broader studies linking these behavioral trends to cognitive maturation are needed.

This gap is particularly concerning given the potential for AI to fundamentally alter cognitive development trajectories, especially among younger users who may be more susceptible to developing dependency patterns. Understanding these long-term effects is crucial for developing educational strategies and AI design principles that protect and enhance human cognitive development. Studies such as Wang and Fan (2025) and Budiyono (2025) have begun to address these concerns, demonstrating that while AI can assist learning, it may also lead to reduced self-regulation and over-reliance without appropriate pedagogical scaffolding. These gaps point to the importance of integrative, longitudinal, and comparative approaches to studying AI and culture. Future research must bridge technical and cultural perspectives, investigating not only how AI adoption is shaped by existing values but also how the presence of AI influences the evolution of those values, priorities, and ethical frameworks within organizations and educational settings (Shneiderman, 2020; Nisbett, 2003; Kasneci et al., 2023). Table 2 summarizes the identified research gaps and connects them with key references.

Table 2. Research Gap, Description and key references

| Research Gap | Description | Key References |
|---|---|---|
| Static treatment of culture | Culture is often treated as fixed rather than dynamic, overlooking reciprocal evolution with AI. | Benbya et al., 2020; Kasneci et al., 2023; Peláez-Sánchez et al., 2024 |
| Limited research on AI and Bias | Culture is often treated as fixed rather than dynamic, overlooking reciprocal evolution with AI; biases in AI risk reinforcing stereotypes and reshaping cultural norms. | Noble, 2018; Benbya et al., 2020; Crawford, 2021; Kasneci et al., 2023; Varsha 2023; Peláez-Sánchez et al., 2024 |
| Methodological limitations (cross-sectional designs) | Over-reliance on cross-sectional studies; lack of longitudinal evidence on trust, adaptation, and collaboration with AI. | Marangunić & Granić, 2015; Prahl & Van Swol, 2021; Fan et al., 2024; Jelson et al., 2025 |
| Limited comparative research | Insufficient exploration of demographic influences (e.g., age, education, language, profession) on AI trust and use. | Zhang & Dafoe, 2019; Schepman & Rodway, 2020; Gerlich, 2025 |
| Narrow focus on culture shaping technology | Focus on how culture shapes AI adoption, with limited attention to how AI reshapes institutional and cultural values. | Shneiderman, 2020; Susskind, 2020; Shen et al., 2023; Milana et al., 2024 |
| Limited research on long-term cognitive effects | Scarce longitudinal studies on AI's impact on cognitive development; concerns over cognitive offloading and critical thinking decline. | Stadler, Bannert & Sailer, 2024; Bai, Liu & Su, 2023; Wang & Fan, 2025; Budiyono, 2025 |
| Need for integrative, longitudinal, and comparative approaches | Calls for approaches that integrate cultural, ethical, and technical dimensions, studying bidirectional dynamics of AI and culture. | Shneiderman, 2020; Nisbett, 2003; Kasneci et al., 2023 |

Source: Authors own illustration

## Research Questions and Hypotheses: AI Integration Dynamics in Academic and Professional Contexts

To reach the research objectives and considering the research gaps identified in the literature review (see Table 2), this investigation is guided by four primary research questions that frame the comparative analysis (Table 3).

**Research Question 1:** *How do trust patterns toward AI technologies differ between academic and business environments, and how do these patterns evolve over time?*

This question explores the foundational element of trust formation, recognizing that academic and business cultures may exhibit different levels of risk tolerance, evaluation criteria, and adoption timelines for AI technologies. Trust in AI has been identified as a crucial regulator of adoption, with increased trust accelerating diffusion and distrust potentially slowing or blocking integration (Glikson & Woolley, 2020; Schoorman, Mayer, & Davis, 2007). Multidisciplinary research on trust in human-AI teams has demonstrated that mechanisms such as institutional endorsement, perceived competence, transparency, and cultural alignment play significant roles in shaping trust trajectories in organizational contexts (Benbya, Nan, Tanriverdi, & Yoo, 2020; Riegelsberger, Sasse, & McCarthy, 2005).

**Research Question 2:** *How do cultural norms and values within academic and business environments influence the perception and integration of AI technologies, and what temporal patterns emerge in this cultural adaptation process?*

This question addresses the deeper cultural mechanisms that shape AI adoption and resistance, examining how established values, expectations, and institutional practices either facilitate or inhibit technological integration. The adoption and spread of technological innovations, including AI, are significantly affected by psychological, cultural, and policy factors, highlighting the need for cultural context analysis when investigating organizational change (Hofstede, Hofstede, & Minkov, 2010; Nisbett, 2003; Scott, 2014).

**Research Question 3:** *What are the differential impacts of AI integration on human problem-solving methodologies and decision-making processes within academic versus business contexts?*

This question investigates the functional outcomes of AI adoption, exploring whether and how different organizational cultures leverage AI technologies in distinct ways and whether these approaches change over time. Research on technology adoption and organizational culture has shown that successful integration requires alignment of people, processes, technology, and institutional readiness (Marangunić & Granić, 2015; Benbya et al., 2020). Furthermore, studies suggest that while AI can enhance efficiency and engagement, it may also introduce new forms of stress or job insecurity, which are experienced differently across organizational environments (Susskind, 2020; Ransbotham, Kiron, Gerbert, & Reeves, 2021).

**Research Question 4:** *Does cognitive dependency on AI technology exist, how do patterns of cognitive dependency on AI technologies differ between academic and business environments, and how do these patterns evolve over time?*

This research question addresses a notable gap in existing literature by focusing on cognitive dependency patterns, recognizing that academic and business cultures may display distinct cognitive demands, educational practices, and acceptance criteria for AI integration. Cognitive dependency on AI has emerged as a critical concern due to its potential to diminish autonomous analytical skills, with increased dependency potentially compromising cognitive growth and problem-solving effectiveness (Gerlich, 2025; Risko & Gilbert, 2016). Prior multidisciplinary research highlights factors such as educational methods, institutional expectations, and cultural norms that shape cognitive interactions with AI (Marangunić & Granić, 2015; Brynjolfsson & McAfee, 2014). A detailed exploration of these differences over time will offer deeper insights into the evolving nature of AI integration and its impacts across diverse contexts.

**Research Methodology**

The formulation of an appropriate methodological framework is essential to ensure systematic and rigorous examination of the research questions guiding this study. Since the research questions constitute the conceptual foundation of the investigation, their accurate answering requires the application of suitable methodological approaches. Table 3 delineates the correspondence between each research question and its associated methodological strategy, thereby providing an integrated roadmap for the empirical inquiry.

Table 3. Research Questions and Corresponding Methods

| Research Question | Methodological Approach |
| --- | --- |
| How do trust patterns differ between academic and business environments? | Longitudinal survey analysis; cross-group comparisons |
| How do cultural norms shape AI perception and integration? | Cross-sectional and longitudinal analyses; qualitative thematic coding |
| What are the differential impacts of AI on problem-solving? | Comparative analysis of academic vs. business contexts |
| Does cognitive dependency differ across contexts? | Mixed-effects longitudinal modelling; survey and qualitative integration |

Source: Authors own illustration

The methodological design follows a mixed-methods orientation, combining quantitative and qualitative strategies to capture both longitudinal dynamics and cross-sectional patterns. This dual orientation facilitates robust analyses of the phenomena under investigation by enabling the triangulation of survey-based measures, comparative group analyses, and qualitative thematic insights. Such an approach ensures not only statistical validity but also contextual depth in the interpretation of findings.

This section outlines the methodological framework guiding the study. It consists of the *general structure of the research (Figure 3)*, detailing the overall approach and rationale, followed by the *study design and temporal structure*, which define the sequencing of activities and timelines. The composition of *sample groups* is presented alongside a *comparative framework of AI integration* across academic and business contexts (figure 4), highlighting their distinct and overlapping dimensions. Furthermore, the section explains the study *wave-specific data collection procedures, the participant dynamics and longitudinal considerations*, and the *data analysis strategy* employed to ensure robustness and validity of findings. Finally, the discussion is extended to cover methodological considerations and limitations, offering critical reflection on the scope and constraints of the research design.

**General structure of the research**

This study, as presented in Figure 3, employs a longitudinal survey design, structured into three main phases: preparation, execution, and analysis.

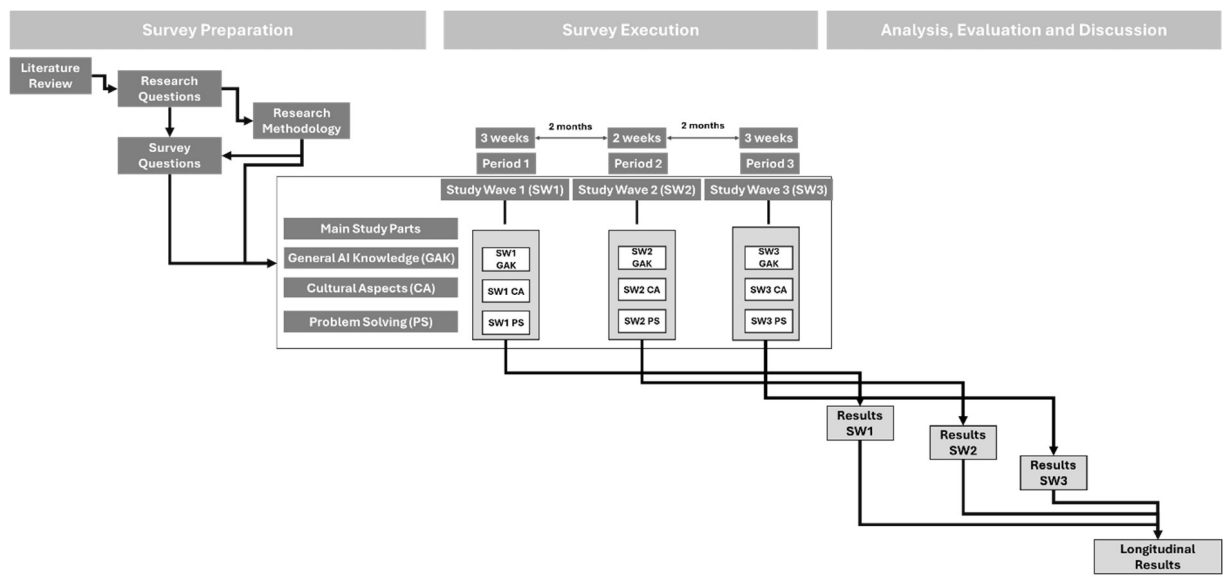

Figure 3: General structure and progression of the longitudinal survey.

Source: Authors own illustration

As shown in Figure 3, the initial phase focused on preparing the study by identifying research gaps through a critical literature review, which informed the choice of methodology and guided the development of the survey instrument. The survey itself was structured into four core sections: (1) demographic information, (2) general AI knowledge, (3) cultural orientations, and (4) problem-solving approaches in AI contexts.

**Study Design and Temporal Structure**

The second phase involved the execution of the longitudinal survey through a three-wave data collection approach (see Figure 3). This design enabled the tracking of participant evolution and the examination of stability and change in perceptions over time. Each wave of the survey was cross-sectional in nature but temporally spaced to capture evolving attitudes, behaviors, and cultural dynamics related to AI usage. The execution of the study is planned as follows:

- Wave 1: January 29th – February 19th, 2025
- Wave 2: April 4th – April 16th, 2025
- Wave 3: June 25th – July 9th, 2025

Participants were recruited through multiple digital channels. LinkedIn was used to reach professionals and researchers, while targeted email invitations were sent to participants affiliated with the European Campus

Rottal-Inn (ECRI) of the Deggendorf Institute of Technology (DIT). For Wave 3, recruitment was extended to include the official DIT Facebook page to further diversify the sample.

**Sample Groups**

Participants were categorized into two primary cohorts based on their professional contexts:

- *Academic group:* Participants affiliated with ECRI, including students, teaching faculty, and administrative staff.
- *Business group:* Professionals unaffiliated with ECRI, primarily from industry and applied research sectors.

This classification facilitates structured comparisons across institutional environments, illuminating how professional contexts distinctly shape AI-related perceptions and behaviours. Within academic settings, trust in AI is often fostered through peer validation, transparency in knowledge exchange, and adherence to ethical norms, emphasizing communal values and scholarly rigor. In contrast, business contexts anchor trust more firmly in return on investment (ROI), operational efficiencies, and mechanisms of corporate accountability, reflecting a pragmatic orientation toward measurable gains. These differences underscore how institutional priorities fundamentally influence the cultivation of trust in AI across sectors.

Recent scholarship substantiates this contextual divergence. For example, a meta-analysis shows that trust in AI is positively correlated with explainability, though mediated by domain-specific expectations (Atf & Lewis, 2025). Similarly, organizational adoption readiness is shaped by experiential engagement with AI's limitations, where sustained trust emerges when peer experiences are formalized into governance frameworks (Übellacker, 2025). At a global scale, survey evidence highlights a persistent trust deficit: despite widespread usage (66%), less than half of respondents (46%) report being willing to trust AI systems, underscoring the salience of institutional and cultural context in trust formation (Gillespie, Lockey, Ward, Macdade, & Hassed, 2025). A Comparative framework of AI integration in Academic Vs Business Contexts is shown under Figure 4.

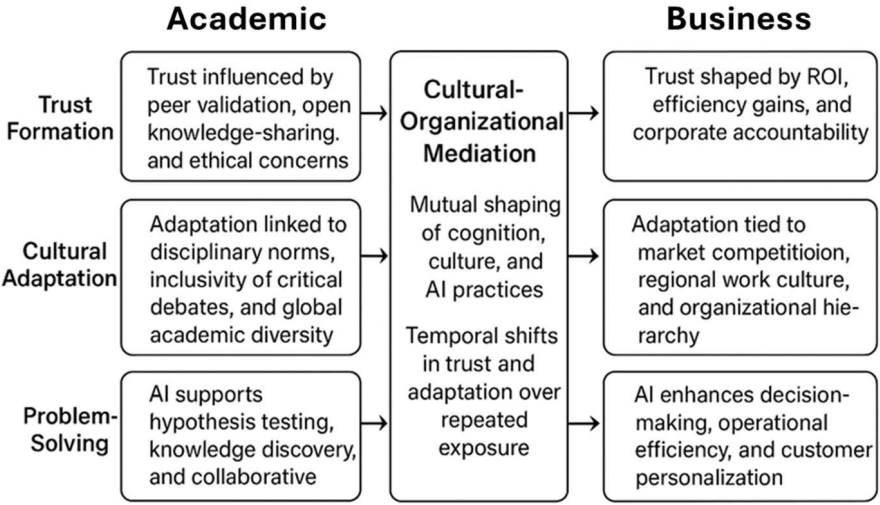

Figure 4: A Comparative framework of AI integration in Academic Vs Business Contexts

Source: Authors own illustration

Figure 4 illustrates the Comparative Framework of AI Integration, contrasting academic and business contexts across three interrelated dimensions: trust formation, cultural adaptation, and problem-solving. In academic settings, AI integration is mediated by values of transparency, ethics, and collaborative inquiry, while in business contexts, adoption is driven by efficiency, competitiveness, and strategic returns. The central mediation layer highlights how cultural and organizational environments shape and are simultaneously reshaped by, AI engagement, leading to evolving norms of trust, cognition, and problem-solving over time. The framework also emphasizes the reciprocal relationship between AI technologies and cultural practices, underscoring how prolonged exposure produces temporal shifts in human behaviour and institutional norms.

### Study Wave-Specific Data Collection

Each wave targeted individuals in knowledge-intensive roles, including advanced students and working professionals with potential or ongoing interaction with AI technologies.

- Wave 1 Protocol: Purposive sampling was conducted through institutional networks, using digital recruitment and stratified sampling to reduce selection bias.
    - *Duration:* 3-week open window
    - *Questionnaire content:* demographics, education, AI background and usage, perceptions of AI's impact on culture, communication, problem-solving, and ethics
    - *Engagement:* A multi-channel recruitment strategy was employed to ensure diverse representation from both academic and business contexts. Email and LinkedIn reminders; option to register for follow-up

- Wave 2 Protocol: For the second wave a stratified digital recruitment was conducted, which was initiated eight weeks after Wave 1 concluded.
    - *Duration:* approx. 2-week data collection window
    - *Questionnaire content:* Questionnaire was identical to Wave 1
    - *Engagement:* A Same strategy as Wave 1, with additional reinforcement via academic lectures and faculty outreach.

- Wave 3 Protocol: In the third wave the digital recruitment with stratified sampling was continued. Wave 3 was started 8 weeks after Wave 2 closure.
    - *Duration:* approx. 3-week data collection window
    - *Questionnaire content:* Questionnaire was identical to Wave 1 and 2
    - *Engagement:* Recruitment channels expanded to include the DIT Facebook page, along with ongoing use of LinkedIn, email, and academic lectures.

### Participant Dynamics and Longitudinal Considerations

The study design accommodates variation in participant engagement across waves by categorizing respondents into three dynamic groups:

- Continuous participants: Individuals completing all three waves, providing rich longitudinal data.
- Intermittent participants: Those participating in non-consecutive waves, offering valuable but partial longitudinal insights.
- Wave-specific participants: New entrants in later waves, contributing cross-sectional data for contemporary attitudes.

This flexible design maximizes data utility across temporal dimensions while enabling robust analysis of both individual- and group-level change over time.

### Data Analysis Strategy

The analytical approach employed in this study utilizes a comprehensive multi-method analysis framework, as depicted in Figure 3, integrating cross-sectional, longitudinal, and mixed-methods approaches to address the research questions comprehensively. This design aligns with established best practices in longitudinal and cross-cultural technology adoption research while incorporating methodological advances in mixed-effects and mixed-methods modeling (Creswell & Plano Clark, 2017; Raudenbush & Bryk, 2002). The goal is to systematically assess AI adoption patterns, trust formation, and changes in problem-solving behavior across time and organizational contexts.

### Cross-Sectional Analysis Framework

Cross-sectional analyses were conducted within each study wave to examine group-specific differences and distributional patterns at discrete time points. Each study wave (SW1, SW2, and SW3) was analyzed independently, with particular focus on three research domains: general AI knowledge, cultural aspects, and problem-solving behaviors. Within each domain, academic and business participant groups were analyzed both separately and comparatively to identify sector-specific patterns and inter-group differences.

Independent-sample t-tests will be employed to compare mean scores between academic and business participants on key outcome variables. To address the potential for Type I error inflation due to multiple comparisons, corrections were applied using the Benjamini-Hochberg procedure (Benjamini & Hochberg, 1995). For categorical data, chi-square tests of independence were utilized to evaluate group differences, while Mann-Whitney U tests served as non-parametric alternatives for analyzing ordinal outcomes that violated normality assumptions.

Cross-domain analyses were conducted within each wave to identify overarching patterns and interactions across the three research areas. This approach facilitated the recognition of emergent themes and relationships between AI knowledge, cultural factors, and problem-solving approaches. Trends across waves were identified through comparative assessments of means, variances, and effect sizes, enabling the detection of evolving patterns in AI adoption and trust formation.

**Longitudinal Analysis Framework**

The longitudinal analysis examined temporal changes across the three study waves for each research domain, both individually and in combination. This analysis focused on evaluating shifts at the group level, comparing trajectories between academic and business participants over the entire data collection period. The longitudinal approach was essential for understanding the dynamic nature of AI adoption and trust development processes.

Descriptive statistics were computed to summarize change patterns and trajectory slopes, highlighting group-specific temporal differences in AI adoption rates, trust levels, and problem-solving strategy evolution. Paired-sample t-tests were conducted to determine statistically significant changes in key outcomes between consecutive waves. Effect sizes were quantified using Cohen's d with 95% confidence intervals to assess both statistical significance and practical relevance of observed changes (Cohen, 1988).

Predictors of significant temporal changes were identified through multiple regression modeling that incorporated demographic variables, baseline characteristics, and contextual influences as covariates. Interaction terms, such as *culture × trajectory* type, were included to test the moderating effect of cultural context on temporal dynamics in AI adoption and trust development. This approach aligns with established findings from cross-cultural technology adoption literature (Nisbett, 2003).

Additionally, hierarchical linear models (multilevel models) were implemented to accommodate the nested data structure arising from repeated observations within individuals across study waves. This analytical approach recognizes that observations within individuals are likely to be more similar to each other than to observations from different individuals, thereby addressing potential violations of independence assumptions inherent in traditional analytical approaches. The mixed-effects models incorporated fixed effects for time (study wave), group membership (academic vs. business), cultural indicators, and theoretically relevant interaction terms. Random effects for both intercepts and slopes were included to account for individual variability in baseline levels and rates of change over time. This specification allows for the examination of both population-level trends and individual differences in change trajectories (Raudenbush & Bryk, 2002).

Model estimation was conducted using restricted maximum likelihood (REML) estimation, selected for its robustness in handling missing data patterns commonly encountered in longitudinal research due to participant attrition. REML provides unbiased parameter estimates under the missing at random (MAR) assumption while maximizing the utilization of available data, thereby preserving statistical power and reducing potential bias from listwise deletion approaches (Little & Rubin, 2019).

**Qualitative Integration Framework**

The mixed-methods component integrates qualitative data through a systematic coding framework applied to open-ended survey responses and any additional qualitative data collected during the study. The coding process focuses on identifying themes related to trust formation mechanisms, concerns about AI misuse, adaptation strategies across organizational contexts, and factors influencing adoption decisions. Thematic analysis follows established procedures for qualitative data analysis, including initial coding, pattern identification, theme development, and theme refinement. Cross-case comparisons explore differences between participants exhibiting significant change patterns and those demonstrating stable response patterns across study waves. This comparative approach enables the identification of factors that distinguish different adoption and trust development trajectories.

Joint display matrices are constructed to systematically link quantitative findings with emergent qualitative themes, following established procedures for mixed-methods integration (Creswell & Plano Clark, 2017). These matrices facilitate the identification of convergent, divergent, and complementary findings between quantitative and qualitative data sources. This integrated approach enables a comprehensive interpretation of results by combining the breadth and generalizability of quantitative analysis with the depth and contextual understanding provided by qualitative insights. The qualitative integration also serves to illuminate potential mechanisms underlying quantitative patterns, providing explanatory depth to observed statistical relationships and supporting the development of more nuanced theoretical models of AI adoption and trust development in organizational contexts.

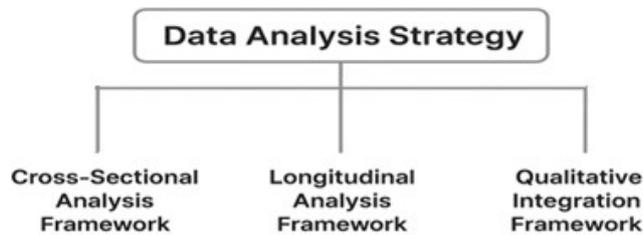

Figure 5: Summary data analysis strategy

Source: Authors own illustration

Figure 5 shows that the study employs a comprehensive multi-method analysis framework that integrates cross-sectional, longitudinal, and qualitative analysis to capture both group-specific patterns and temporal dynamics in AI adoption and trust development. Cross-sectional analyses highlight inter-group and domain-specific differences at discrete time points, while longitudinal models trace change trajectories and account for individual variability over time. The qualitative integration adds explanatory depth by linking emergent themes to quantitative findings, enabling a holistic understanding of adoption processes across cultural and organizational contexts.

**Methodological Considerations and Limitations**

The dynamic nature of the participant pool presents several methodological challenges that must be addressed in the analysis and interpretation of results. The combination of longitudinal and cross-sectional data requires careful statistical treatment to ensure the validity of inferences, particularly regarding patterns of change over time versus group differences at single time points. Hierarchical modeling and robust data linkage procedures are employed to mitigate these complexities, consistent with best practices in longitudinal research (Raudenbush & Bryk, 2002). The recruitment strategy's reliance on multiple channels, including email outreach, in-person presentations, and professional networks, may introduce selection biases that could affect the generalizability of findings. For example, individuals more engaged with digital platforms or already interested in AI may be overrepresented in the sample. Efforts to diversify recruitment and track attrition are intended to address this limitation, but some bias may persist, as is often the case in mixed-methods research (Creswell & Plano Clark, 2017).

The temporal spacing of waves (about 8-week pause between the waves) was chosen to balance the need for meaningful change observation with practical considerations of participant retention. However, this timeframe may not fully capture longer-term dynamics in AI adoption or more gradual processes of cultural adaptation. As such, the study's conclusions should be interpreted within these temporal boundaries, a common consideration in studies of organizational change (Scott, 2014). Despite these limitations, the mixed-methods and multi-wave design strengthen the robustness and richness of the findings, while acknowledging the inherent trade-offs present in real-world organizational research (Creswell & Plano Clark, 2017; Little & Rubin, 2019).

**Expected Contributions and Implications**

This study offers substantial theoretical and practical advancements in understanding the interplay between AI and human culture, while also addressing the cognitive transformations occurring in response to AI use. By integrating sociocultural, organizational, and cognitive perspectives, the findings provide a multidimensional foundation for designing responsible AI systems that align with human capacities and cultural frameworks.

### Theoretical Contributions

The investigation advances theoretical understanding across four interconnected domains, extending existing frameworks and contributing novel empirical evidence to the literature.

### Cultural Dimensions Integration in Technology Adoption Theory

The present research extends established technology adoption frameworks, including the TAM and Unified Theory of Acceptance and Use of Technology, through systematic incorporation of cultural dimensions as theoretical constructs. Specifically, the investigation integrates societal normative structures, value orientation systems, and trust architecture as moderating variables influencing artificial intelligence adoption trajectories. Empirical findings demonstrate that organizational contextual factors exert significant conditioning effects on implementation pathways and effectiveness outcomes, thereby necessitating theoretical model extension to accommodate cultural variability as a fundamental parameter rather than peripheral consideration (Hofstede et al., 2010; Marangunić & Granić, 2015; Nisbett, 2003).

### Bidirectional AI-Culture Interaction Dynamics

This study contributes empirical evidence supporting the reciprocal nature of artificial intelligence adoption and organizational cultural evolution processes. Contrary to technological determinism paradigms that position technology as a unidirectional change agent, the findings reveal mutual adaptation mechanisms wherein artificial intelligence systems simultaneously influence and respond to cultural normative structures, value systems, and organizational practices. This bidirectional relationship necessitates theoretical models capable of capturing dynamic interactions between technological and cultural subsystems rather than linear causation frameworks (Benbya et al., 2020; Scott, 2014).

### Temporal Patterns in AI Acculturation and Cognitive Adaptation

Through longitudinal analytical approaches, this research identifies systematic phase-based integration patterns characterizing artificial intelligence incorporation into professional practice environments. Temporal data analysis reveals distinct stages encompassing normative internalization processes, trust recalibration mechanisms, and adaptive behavioral modification sequences. These findings contribute novel theoretical insights into AI acculturation processes within knowledge-intensive organizational contexts, documenting previously uncharacterized temporal dynamics governing artificial intelligence embedding within established organizational routines and individual workflow architectures (Raudenbush & Bryk, 2002).

### AI-Human Cognitive Interface Theoretical Framework

The investigation contributes to emerging theoretical understanding of artificial intelligence effects on human cognitive function through examination of cognitive offloading phenomena, attentional resource redistribution, and mental workload modulation mechanisms. Empirical analysis identifies both facilitative and inhibitive impacts of artificial intelligence support on human problem-solving capacity. While artificial intelligence demonstrates scaffolding potential for higher-order reasoning under temporal or knowledge constraints, evidence indicates potential erosion of independent analytical capabilities under conditions of excessive utilization or uncritical dependency. These findings inform theoretical conceptualization of artificial intelligence as a cognitive partner engaged in co-constitutive thinking process modification (2024; Ahmad et al., 2023; Ward, 2013).

### Practical Implications

Beyond its theoretical contributions, this investigation yields substantive practical implications across three domains: organizational strategy formulation, cross-cultural artificial intelligence deployment protocols, and human-centered technology design frameworks.

### Culturally Adaptive Artificial Intelligence Implementation

The present study advances implementation protocols incorporating cultural sensitivity as a foundational element in artificial intelligence system architecture and deployment strategies. Evidence suggests that artificial intelligence applications calibrated to organizational cultural values and communication paradigms demonstrate enhanced system legitimacy, elevated user engagement metrics, and improved ethical alignment outcomes (Creswell & Plano Clark, 2017). This approach addresses the documented gap between technological capability and contextual appropriateness in organizational settings.

### Cross-Cultural Integration Framework for Conflict Mitigation

A comprehensive cross-cultural artificial intelligence integration framework emerges from this research to facilitate global deployment initiatives. The framework enables systematic diagnosis of cultural friction sources, including misaligned trust normative structures and divergent perspectives on human-machine collaborative arrangements. Through identification of these friction points, the framework supports development of adaptive strategies that enhance inclusivity and minimize resistance across geographical and cultural boundaries (Meyer, 2014; Hall, 1976).

### Professional Development Pathways for AI-Augmented Cognitive Performance

Drawing from observed cognitive transformation patterns, this study delineates structured professional development pathways that synthesize artificial intelligence literacy with metacognitive skill enhancement and trust calibration protocols. These developmental frameworks target adaptive expertise cultivation, enabling practitioners to optimize artificial intelligence capabilities while preserving critical analytical thinking and domain-specific problem-solving competencies (Holmes, Bialik, & Fadel, 2019; Shneiderman, 2020; Norman, 1993).

**Mitigation Strategies for Cognitive Dependency Risk**

The findings illuminate the phenomenon of potential cognitive deskilling resulting from excessive artificial intelligence dependency. Evidence-based recommendations emerge for promoting balanced artificial intelligence utilization, including implementation of alternating assisted and independent problem-solving modalities. These approaches sustain human agency and reflective cognitive capacity within AI-enhanced operational environments (2024; Kirsch, 2021).

**Conclusion**

This work introduces a rigorous methodological and theoretical framework for examining the dynamic interplay between artificial intelligence (AI), human cognition, problem-solving, and cultural adaptation in academic and business environments. Employing a three-wave longitudinal design over five months, the framework captures temporal dynamics (such as evolving trust, perceived AI competence, and cultural adaptation) that are typically overlooked in cross-sectional studies. By integrating a dynamic participant pool reflecting real-world organizational fluidity, the methodology enables observation of both sustained and transient user behaviours during AI integration. Guided by three core research questions (how individuals perceive AI competence, how cultural differences shape AI adoption, and how AI transforms cognitive and problem-solving processes) the study combines temporal, cultural, and cognitive dimensions into a bidirectional framework. This approach demonstrates that AI adoption is not merely a response to pre-existing cultural contexts but actively interacts with and reshapes organizational norms and cognitive engagement over time, challenging deterministic models of technological change.

      The methodological contribution lies in linking longitudinal observation with culturally and cognitively informed metrics, allowing for fine-grained analysis of adoption trajectories, trust formation, and problem-solving evolution. This framework supports the design of human-centred, culturally responsive AI strategies by highlighting context-specific adoption patterns, temporal trust development, and adaptive cognitive engagement across organizational settings. Comparative insights reveal that academic environments foster collaborative and deliberative AI practices, whereas business contexts prioritize efficiency and performance, illustrating the necessity of context-sensitive implementation strategies. By emphasizing temporal dynamics, cultural sensitivity, and cognitive outcomes, this methodological blueprint provides a replicable foundation for future research on human-AI co-evolution. It supports both quantitative and qualitative extensions, offering pathways to uncover deeper mechanisms of norm change, trust formation, and cognitive transformation in organizational contexts. Ultimately, this study advances methodological rigor in AI research, demonstrating that meaningful AI integration requires attention to cognitive, cultural, and organizational dimensions. The framework ensures that AI adoption aligns with ethical values, social practices, and sustainable human development, providing a robust tool for investigating and guiding the evolving landscape of human-AI interaction.